\documentclass[a4paper]{article}

\usepackage[utf8]{inputenc}
\usepackage[top=10mm]{geometry}
\usepackage{booktabs}
\usepackage[flushleft]{threeparttable}
\usepackage{authblk}
\usepackage{bm}
\usepackage{amsmath}
\usepackage{amssymb}
\usepackage{graphicx}
\usepackage[square,sort,numbers]{natbib}
\usepackage{slashed}
\usepackage{float}
\usepackage{longtable,booktabs}
 \allowdisplaybreaks[4]

\title{Aspects of $Z_{cs}(3985)$ and $Z_{cs}(4000)$}

\author[1]{Shuai Han}
\author[1,2]{Li-Ye Xiao \thanks{E-mail: lyxiao@ustb.edu.cn}}
\affil[1]{\textit{\small Institute of Theoretical Physics, University of Science and Technology Beijing,
Beijing 100083, China}}
\affil[2]{\textit{\small Department of
Physics, Hunan Normal University, and Key Laboratory of
Low-Dimensional Quantum Structures and Quantum Control of Ministry
of Education, Changsha 410081, China}}

\date{\today}

\begin{document}
\maketitle

%abstract
\begin{abstract}
In the present work we investigate the $\eta_c K$, $J/\psi K$, $\eta_c K^*$ and $J/\psi K^*$ hidden-charm decay modes for the $c\bar{c}s\bar{u}$ four-quark system in the molecular and compact tetraquark scenarios using the quark-exchange model. Our theoretical results indicate that if the newly observed states $Z_{cs}(3985)$ and $Z_{cs}(4000)$ are two different states, $Z_{cs}(4000)$ may be interpreted as the mixture $\frac{1}{\sqrt{2}}(D^0D_s^{*-}+D^{*0}D_s^{-})$ of which the $J/\psi K$ partial decay width is about $\Gamma\sim2.89$ MeV, while $Z_{cs}(3985)$ may be explained as the mixture $\frac{1}{\sqrt{2}}(-D^0D_s^{*-}+D^{*0}D_s^{-})$ of which the $J/\psi K$ partial decay width is small to zero.
Moreover, if the state $Z_{cs}(4000)$ can be explained as the mixed state $\frac{1}{\sqrt{2}}(D^0D_s^{*-}+D^{*0}D_s^{-})$ indeed, the partial decay width ratio between $J/\psi K$ and $\eta_cK^*$ is close to unit, which indicates the decay channel $\eta_cK^*$ may be a ideal channel as well to decode the inner structure of $Z_{cs}(4000)$. In addition, the partial decay width for the tensor molecular state $|D^{*0}D_s^{*-}\rangle_{2^+}$ decaying into $J/\psi K^*$ can reach up to a few MeV, which shows this tensor molecular state has a good potential to be observed in this decay channel.

\end{abstract}

\section{Introduction}
In the last twenty years there is an explosion in the observation of hidden-charmed multiquark states~\cite{Esposito:2016noz,Brambilla:2019esw,Liu:2019zoy,Olsen:2017bmm,Guo:2017jvc,Chen:2016qju}. For example: dozens of charmonium-like states have been observed since 2003~\cite{Tanabashi:2018oca}, most of which are good candidates of four-quark states with quark components $c\bar{c}q\bar{q}$($q$ denotes $u$ or $d$ quark)\cite{Esposito:2016noz,Brambilla:2019esw,Liu:2019zoy,Olsen:2017bmm,Guo:2017jvc,Chen:2016qju}; The LHCb collaboration reported hidden-charm pentaquark states $P_c(4380)$ and $P_c(4450)$ in 2015~\cite{Aaij:2015tga} and updated the data in 2019~\cite{Aaij:2019vzc}, which minimally contain four quarks and one antiquark($c\bar{c}qq\bar{q}$)~\cite{Zhang:2019lbg}; Last year, the LHCb collaboration observed a narrow structure around 6.9 Gev in the $J/\psi$-pair invariant mass spectrum~\cite{Aaij:2020fnh}, which may be good candidates of compact tetraquark state $cc\bar{c}\bar{c}$~\cite{Chao:2020dml}. All of the experimental progress in the observation of hadrons with anomalous properties has attracted a great deal of attention from the hadron physics community.

Very recently the BESIII collaboration made a great
breakthrough in searching for hidden-charmed multiquark
states with strangeness and reported a new structure near the $D^0D_s^{*-}$ and $D^{*0}D_s^{-}$ mass thresholds in the $K^+$ recoil-mass spectrum~\cite{Ablikim:2020hsk}, whose mass and width are respectively
\begin{eqnarray}
M[Z_{cs}(3985)]&=&3982.5^{+1.8}_{-2.6}\pm2.1~\text{MeV},\nonumber\\
\Gamma[Z_{cs}(3985)]&=&12.8^{+5.3}_{-4.4}\pm3.0~\text{MeV}.
\end{eqnarray}
From its decay modes, it is reasonable to assign it as the first candidate of the hidden-charm tetraquark with strangeness($c\bar{c}s\bar{q}$). Thus, Most of theoretically references are under the four-quark scenario to exploring its genuine properties of the excess~\cite{Meng:2020ihj,Wang:2020htx,Simonov:2020ozp,Xu:2020evn,Sungu:2020zvk,Guo:2020vmu,Wang:2020dgr,Dong:2021juy,Ozdem:2021yvo,Yan:2021tcp,
Zhu:2021vtd,Wan:2020oxt,Du:2020vwb,Chen:2020yvq,Sun:2020hjw,Wang:2020rcx,Cao:2020cfx,Azizi:2020zyq,Wang:2020iqt,Jin:2020yjn,Yang:2020nrt,Liu:2020nge,Albuquerque:2021tqd}.
There also exist a few discussions that assign this structure as a threshold effect~\cite{Ikeno:2021ptx} or a threshold cusp~\cite{Dong:2020hxe}.
Furthermore, explaining $Z_{cs}(3985)$ as a reflection structure of charmed-strange meson $D_{s2}(2573)$ is also presented~\cite{Wang:2020kej}.

Later the LHCb collaboration further observed a exotic states with quark content $c\bar{c}u\bar{s}$(denoted as $Z_{cs}(4000)$) from an amplitude analysis of the $B^+\rightarrow J/\psi \phi K^+$ decay~\cite{Aaij:2021ivw}. Its mass and width are measured to be
\begin{eqnarray}
M[Z_{cs}(4000)]&=&4003\pm6^{+4}_{-14}~\text{MeV},\nonumber\\
\Gamma[Z_{cs}(4000)]&=&131\pm15\pm26~\text{MeV}.
\end{eqnarray}
The mass of this state is comparable to that of $Z_{cs}(3985)$ observed by BESIII~\cite{Ablikim:2020hsk}, while its width is about ten times larger than that of $Z_{cs}(3985)$. Thus, whether they are two different states or not, and how to decode their inner structures are now pretty much the agenda for theorists. So far, there are some discussions of their inner structures with various theoretical methods~\cite{Shi:2021jyr,Ortega:2021enc,Chen:2021erj,Maiani:2021tri,Ge:2021sdq,Giron:2021sla,Ozdem:2021hka,Meng:2021rdg,Yang:2021zhe,Karliner:2021qok}. The most papular interpretation of $Z_{cs}(3985)/Z_{cs}(4000)$ is explaining them as two different states: $DD_s^*/D^*D_s$ molecular states~\cite{Chen:2021erj,Meng:2021rdg} or compacted tetraquark states $c\bar{c}s\bar{u}$ with different spin-parity~\cite{Maiani:2021tri,Shi:2021jyr,Yang:2021zhe}.

The masses of the $Z_{cs}(3985)$ and $Z_{cs}(4000)$ states are comparable and both slightly higher than the mass thresholds of $DD_s^*$ and $D^*D_s$. In the molecular scenario, we take the $Z_{cs}(3985)$ and $Z_{cs}(4000)$ states as two $DD_s^*$/$D^*D_s$ resonance molecular states. Meanwhile, according to experimental measurement, the $Z_{cs}(4000)$ state is observed in the $J/\psi K$ decay channel and non-observation of the $Z_{cs}(3985)$ signal in
this decay mode. Thus, to clarify whether $Z_{cs}(3985)$ and $Z_{cs}(4000)$ are two different states and further show light on the inner structures of the two states, it is critical to study the $J/\psi K$ decay mode for the $DD_s^*$ and $D^*D_s$ molecular states. In addition, the authors in Ref.~\cite{Meng:2021rdg} predicted the existence of a tensor $D^*D_s^*$ resonance. As the expansion it is necessary to investigate the decay properties of this tensor resonance.  Based on the above considerations, in the present work we conduct a systematically study of the hidden-charm strong decays of the $D^{(*)}D_s^{(*)}$ molecular states in the framework of the quark-exchange model. Furthermore, for comparison we also investigate the hidden-charm decay properties of the four-quark system  $c\bar{c}s\bar{u}$ in the compact tetraquark scenario. We hope that our theoretical results can provide some reference for the properties of the $Z_{cs}(3985)$ and $Z_{cs}(4000)$ states.

This paper is structured as follows. In Sec.II, we give a brief introduction of the quark-exchange model. In Sec.III, we present our theoretical results and discussions. Finally, we make a short summary in Sec.IV.

\section{Model introduction}\label{model}

In this work we calculate the $J/\psi K$, $\eta_c K$, $\eta_c K^*$ and $J/\psi K^*$ decay modes for the four-quark system $c\bar{c}s\bar{u}$ in the molecular and compact tetraquark scenarios using the quark-exchange model~\cite{Barnes:2000hu}. This phenomenological model has been used to study the hidden-charm decay properties for the milt-quark states in our previous works~\cite{Wang:2019spc,Xiao:2019spy,Wang:2020prk}. Here we give a brief presentation for this model and more detail information can refer to our previous work~\cite{Xiao:2019spy}.

$\mathbf{A.}~\mathbf{Decay~width}$~~
For a two-body decay process($I\rightarrow C+D$), the decay width in the rest frame of the
initial particle reads
%For a four-quark state ($F$, for short) decaying into two particles
%labelled as $C$ and $D$, the decay width in the rest frame of the
%initial particle has the form
\begin{eqnarray}
d\Gamma=\frac{|\vec{p}_c|}{32\pi^2M^2}|\mathcal{M}(I\rightarrow
C+D)|^2d\Omega.
\end{eqnarray}
Here, $\vec{p}_c$ represents the three-momentum of the final meson $C$; $M$ is the mass of the initial state $I$;
$\mathcal{M}(I\rightarrow C+D)$ denotes the transition amplitude, which is related to the
$T$-matrix via
\begin{eqnarray}
\mathcal{M}(F\rightarrow
C+D)=-(2\pi)^{3/2}\sqrt{2M}\sqrt{2E_C}\sqrt{2E_D}T,
\end{eqnarray}
where $E_C$ and $E_D$ represent the energy of the final mesons $C$ and
$D$, respectively. The $T$-matrix has the form
\begin{eqnarray}
T&=&\langle\psi_{CD}(\vec{p}_c)|V_{\text{eff}}(\vec{k},\vec{p_c})|\psi_{I}(\vec{k})\rangle\nonumber\\
&=&\langle\psi_{CD}(\vec{p}_c)|V_{\text{eff}}(\vec{k},\vec{p_c})|\psi_{AB}(\vec{k})\rangle.
\end{eqnarray}
 Here, the initial state is a four-quark state, which concludes constituent clusters $A$ and $B$. In molecular scenario, the constituents are mesons, while in tetraquark scenario the
constituents are the diquark $[cq]$ and antidiquark
$[\bar{c}\bar{q}]$. Thus, $\psi_{AB}(\vec{k})$
is the normalized relative spatial wave function between the
constituent clusters $A$ and $B$. $\psi_{CD}(\vec{p}_c)$ denotes the relative spatial wave
function between the final mesons $C$ and $D$. $V_{\text{eff}}(\vec{k},\vec{p_c})$ represents the
effective potential, which is a function of the
initial and final relative momentum $\vec{k}$ and $\vec{p}_c$.

Considering the four-quark state may be a superposition of terms with different
orbital angular momenta, the relative spatial
wave function in the momentum space reads
\begin{eqnarray}
\psi_{AB}(\vec{k})=\sum_lR_{nl}(k)Y_{lm}(\frac{\vec{k}}{k}).\label{relative}
\end{eqnarray}
Then, the Eq.~(4) can be written as
\begin{eqnarray}
T&=&\frac{1}{(2\pi)^3}\int d\vec{k}\int d\vec{p}\delta(\vec{p}-\vec{p}_c)V_{\text{eff}}(\vec{k},\vec{p_c})\sum_lR_{nl}(k)Y_{lm}(\frac{\vec{k}}{k})\nonumber\\
&=&\frac{1}{(2\pi)^2}\sum_lM_{ll}Y_{lm}(\frac{\vec{p}_c}{p_c}),
\end{eqnarray}
where
\begin{eqnarray}
M_{ll}=\int_{-1}^{1}P_l(\mu)d\mu \int
dkV_{\text{eff}}(\vec{k},\vec{p}_c,\mu)R_{nl}(k)k^2.
\end{eqnarray}
In this equation, $\mu$
represents the cosine of the angle between the
momenta $\vec{k}$ and $\vec{p}_c$; $P_l(\mu)$ is Legendre function.

Finally, the decay width of
two-body decay progress with the relativistic phase space has the form
\begin{eqnarray}
\Gamma=\frac{E_CE_D|\vec{p}_c|}{(2\pi)^3M}|M_{ll}|^2.\label{gama1}
\end{eqnarray}

\begin{figure}[t]
\centering
\includegraphics[height=8cm,width=8cm]{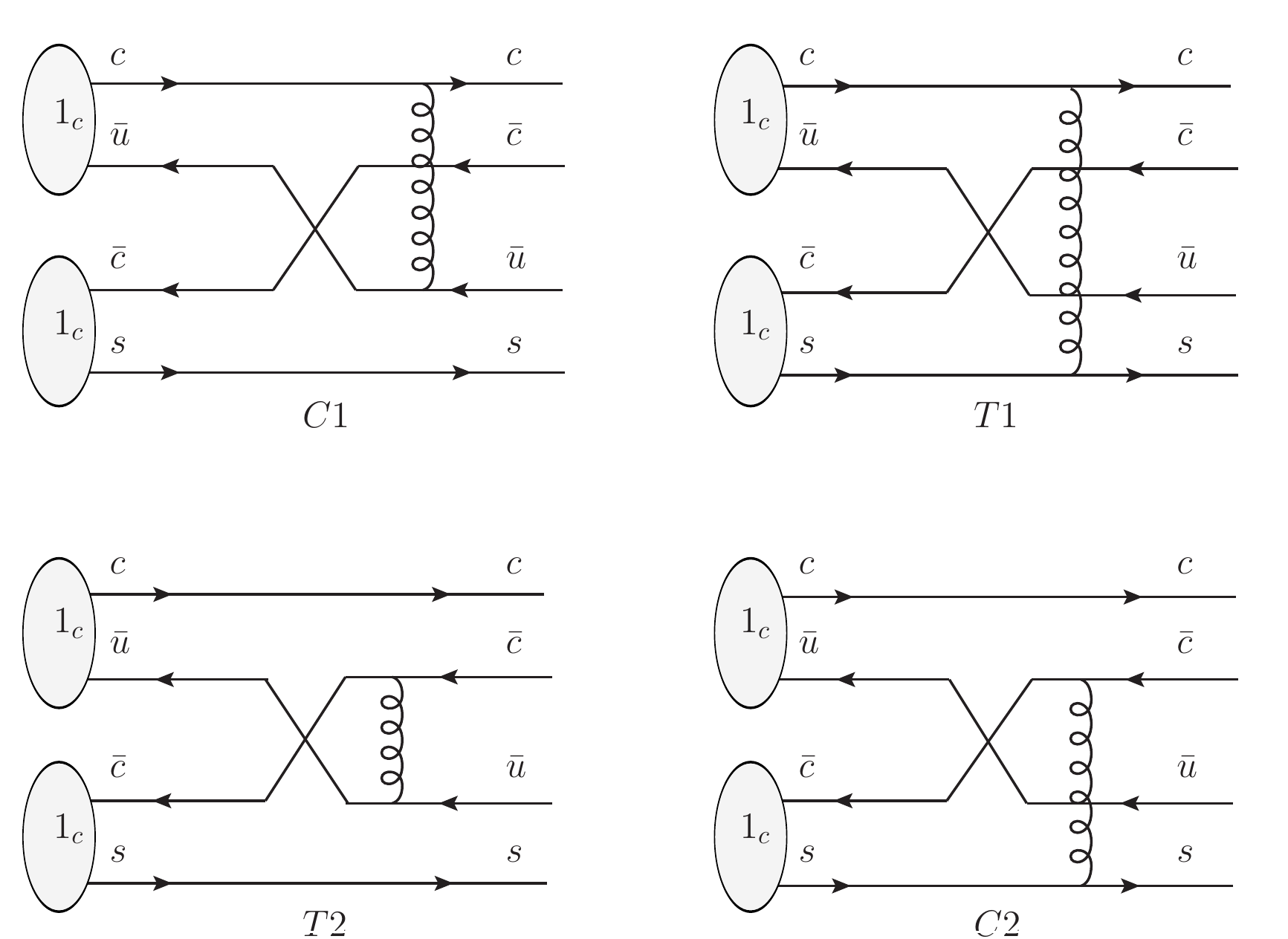}
\caption{Diagrams for
the scattering process $AB\rightarrow CD$ in the molecular scenario.
}\label{post1}
\end{figure}

$\mathbf{B.~Effective~potential}$~~
In the molecular scenario, we treat the four-quark system as
the loosely bound $S$-wave $D^{(*)}D_s^{(*)}$
 molecular states.
% \begin{equation}
% Z_{cs}:~\text{cos}\theta~D^0D_s^{*-}+\text{sin}\theta~D^{*0}D_s^-.
% \end{equation}
 At Born order, the effective potential $V_{\text{eff}}(\vec{k},\vec{p}_c,\mu)$ is related to the reacting amplitude of the meson-meson scattering process, which is
estimated by the sum of the interactions between the inner quarks as
illustrated in Fig.~\ref{post1} within the quark-exchange model. Moreover, the short-range
interactions are dominant and can be
approximated by the one-gluon-exchange (OGE) potential $V_{ij}$ at
quark level,
%\begin{eqnarray}
%A(12)+B(34)\rightarrow C(13)+D(24),
%\end{eqnarray}
%where 1(3) and 2(4) denote the $c$($\bar{c}$) quark and
%$\bar{q}$($q$) quark, respectively. In the quark interchange
%model~\cite{Barnes:1991em,Swanson:1992ec,Hilbert:2007hc,Barnes:1999hs,Barnes:2000hu},
%the scattering Hamiltonian of the processes
%$D^0(D^{*0})+D_s^{*-}(D_s^{-})\rightarrow \eta_c(J/\psi)+K^*(K/K^*)$ is
%estimated by the sum of the interactions between the inner quarks as
%illustrated in Fig.~\ref{post1}. Moreover, the short-range
%interactions are dominant in the scattering processes of two
%open-charmed mesons into a ground charmonium state plus a
%light-flavor meson. Thus, the scattering potential can be
%approximated by the one-gluon-exchange (OGE) potential $V_{ij}$ at
%quark level~\footnote{The interactions in Eq.~(\ref{Q3}) are the
%Fourier transformation of the potential in Ref.~\cite{Wong:2001td}.
%In the following. we perform our calculations in the momentum space
%for the purpose of simplification. The constant potential in the
%spatial space does not contribute due to the cancelation of the form
%factors and we just omitted the term in Eq.~(\ref{Q3}).},
\begin{eqnarray}
V_{ij}=\frac{\lambda_i}{2}\frac{\lambda_j}{2}\left\{\frac{4\pi\alpha_s}{q^2}+\frac{6\pi
b}{q^4}-\frac{8\pi\alpha_s}{3m_im_j}\mathbf{s}_i\cdot\mathbf{s}_je^{-\frac{q^2}{4\sigma^2}}\right\}\label{Q3},
\end{eqnarray}
where $\lambda_i(\lambda^T_i)$ is the quark (antiquark)
generator; $q$ represents the transferred momentum; $b$ corresponds to the string
tension;  $m_i~(m_j)$ and $\mathbf{s}_i~(\mathbf{s}_j)$
are the interacting constituent quark mass and spin
operator; $\sigma$ denotes the range parameter in the hyperfine spin-spin
interaction; $\alpha_s$ represents the running coupling constant,
\begin{eqnarray}
\alpha_s(Q^2)=\frac{12\pi}{(33-2n_f)\text{ln}(A+Q^2/B^2)}\label{Q4}.
\end{eqnarray}
Here, $Q^2$ denotes the square of the invariant masses of the
interacting quarks.
%The parameters in Eqs.~(\ref{Q3})-(\ref{Q4}) are
%fitted by the mass spectra of the observed
%mesons~\cite{Wong:2001td}, and their values are listed in
%Table~\ref{parameters}.

%\begin{table}[h]
%\caption{\label{parameters} The parameters~\cite{Wong:2001td} used
%in the quark model.}
%\begin{tabular}{lllcccccccccccl}\hline\hline
%Parameter~~&$b$~~~~~~~~&0.18~~~~GeV$^2$\\
%         ~~&$\sigma$~~~~~~~~&0.897~~GeV\\
%         ~~&$A$~~~~~~~~&10\\
%        ~~&$B$~~~~~~~~&0.31~~~~GeV\\
%\hline
%Constituent quark mass~~&$m_q$~~~~~~~~&0.334~~GeV\\
%                      ~~&$m_s$~~~~~~~~&0.575~~GeV\\
%                      ~~&$m_c$~~~~~~~~&1.776~~GeV\\
%\hline\hline
%\end{tabular}
%\end{table}

In the quark model, the wave function for a
meson is
\begin{eqnarray}
\Psi=\omega_c\phi_f\chi_s\psi(\vec{p}),
\end{eqnarray}
where $\omega_c$, $\phi_f$, $\chi_s$ and $\psi(\vec{p})$ represent
the wave functions in the color, flavor, spin and momentum space,
respectively.
Thus, the effective potential can be given as the product of the factors,
\begin{eqnarray}
V_{\text{eff}}(\vec{k},\vec{p}_c,\mu)=I_{\text{color}}I_{\text{flavor}}I_{\text{spin-space}}.\label{effective}
\end{eqnarray}
In this equation, $I$ with the subscripts color, flavor and spin-space represent
the overlaps of the initial and final wave functions in the
corresponding space.

In addition, to calculate the decay
widths by Eq.~(\ref{gama1}), expecting the obtained effective potential
$V_{\text{eff}}(\vec{k},\vec{p}_c,\mu)$, we also need the
relative spatial wave function $\psi_{AB}(\vec{k})$ between mesons
$A$ and $B$. In this work, we adopt an $S$-wave harmonic
oscillator function to estimate the
relative spatial wave function in Eq.~(\ref{relative}), which is
\begin{eqnarray}
R_{00}(k)=\frac{2\text{exp}^{-\frac{k^2}{2\alpha^2}}}{\pi^{1/4}\alpha^{3/2}}.\label{relativeMOLECULAR}
\end{eqnarray}
The value of the harmonic oscillator strength $\alpha$ is related to
the root mean square radius $r_{\text{mean}}$, which varies in the range of (1.0-3.0)~fm in the present work.

\begin{figure}[t]
\centering
\includegraphics[height=8cm,width=8cm]{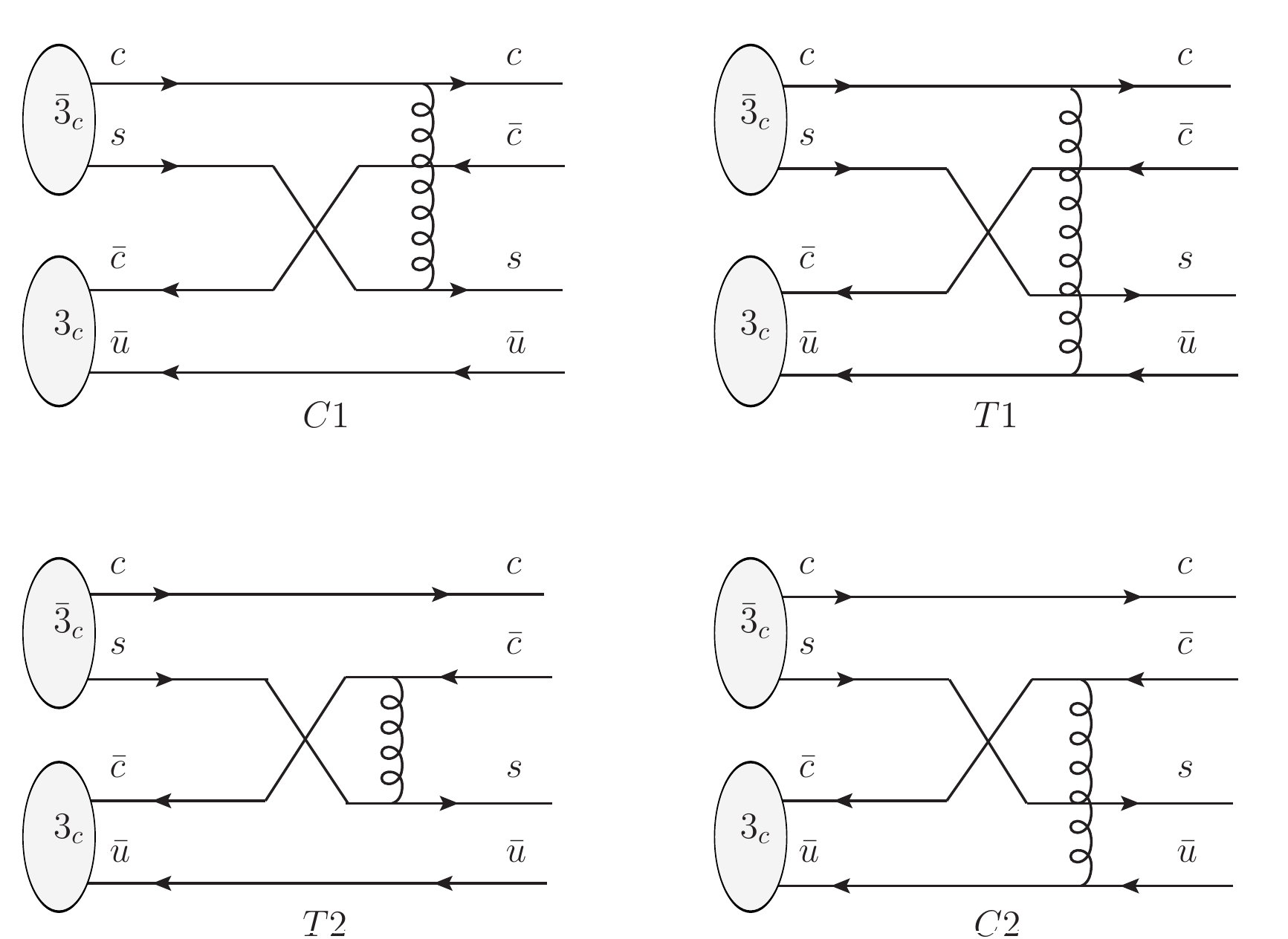} \caption{Diagrams for
the scattering process $AB\rightarrow CD$ in the tetraquark
scenario. }\label{post2}
\end{figure}

Similar to the molecular case, the
$V_{\text{eff}}(\vec{k},\vec{p}_c,\mu)$ in the tetraquark scenario can be approximated by the
interaction between the inner quarks(shown in Fig.~\ref{post2}), and be obtained with Eq.~(\ref{effective})
as well. However there is a difference between the two scenarios for the color factor $I_{\text{color}}$: in the molecular scenario the color
configuration is $1_c$-$1_c$, while in the tetraquark scenario it is $3_c$-$\bar{3}_c$. The difference in color configurations may
result in quite different decay properties.

The spatial wave function of the initial tetraquark states are estimated by the $S$-wave harmonic oscillating wave
function,
\begin{eqnarray}
\Psi(\vec{k}_r,\vec{k}_R,\vec{k}_X)=\psi_A(\vec{k}_r,\alpha_r)\psi_B(\vec{k}_R,\alpha_R)\psi_{AB}(\vec{k}_X,\alpha_X),
\end{eqnarray}
where $\vec{k}_{r/R}$ represent the momentum between the $c(\bar{c})$
and $s(\bar{u})$ quarks in the diquark (antidiquark), and
$\vec{k}_X$ denotes the one between the diquark $[cs]$ and antidiquark
$[\bar{c}\bar{u}]$. The $\alpha$ with the subscripts is the
oscillating parameter along the corresponding Jacobi coordinates, which value is given via imitating the wave function of $Z_{c}(3900)$ in Ref.~\cite{Deng:2015lca} and listed in Table~\ref{tetraquark}.

\begin{table}[h]
\centering\caption{\label{tetraquark} The rms of the tetraquark states
$[cs][\bar{c}\bar{u}]$. $\sqrt{\langle r\rangle^2}$($\sqrt{\langle
R\rangle^2}$) denotes the distance between $c(\bar{c})$ and
$s(\bar{u})$ quarks; $\sqrt{\langle X\rangle^2}$ is the distance
between the diquark $[cs]$ and antidiquark $[\bar{c}\bar{u}]$; unit
of rms is fm.}
\begin{tabular}{lcccccccccccccc}\hline\hline
\text{States}~~&\text{tetraquark}~~&$\sqrt{\langle r\rangle^2}$~~&$\sqrt{\langle R\rangle^2}$~~&$\sqrt{\langle X\rangle^2}$\\
\hline
$Z_c(3900)$~\cite{Deng:2015lca}~~&$\big[[cd]^{S=0}_{\bar{3}_c}[\bar{c}\bar{u}]^{S=1}_{3_c}\big]^{S=1}_{1_c}$~~&0.90~~&0.90~~&0.48\\
$Z_{cs}$~~&$\big[[cs]^{S=0(1)}_{\bar{3}_c}[\bar{c}\bar{u}]^{S=1(0)}_{3_c}\big]^{S=1}_{1_c}$~~&0.90~~&0.90~~&0.48\\
        ~~&$\big[[cs]^{S=1}_{\bar{3}_c}[\bar{c}\bar{u}]^{S=1}_{3_c}\big]^{S=0,1,2}_{1_c}$~~&0.90~~&0.90~~&0.48\\
\hline\hline
\end{tabular}
\end{table}

\section{Results and discussion }\label{results}

The narrow state $Z_{cs}(3985)$ reported by BESIII collaboration~\cite{Ablikim:2020hsk} is observed in the $D^0D_s^{*-}$ and $D^{*0}D_s^-$ final channels, while the wide state $Z_{cs}(4000)$ reported by LHCb collaboration~\cite{Aaij:2021ivw} is observed in the $J/\psi K$ decay mode.  The mass thresholds of $D^0D_s^{*-}$($\sim3977$ MeV) and $D^{*0}D_s^-$($\sim3975$ MeV) are slightly lower than the masses of $Z_{cs}(3985)$ and $Z_{cs}(4000)$. Hence it is vital to investigate the decay properties of the molecular states $D^0D_s^{*-}$ and $D^{*0}D_s^-$. Meanwhile, the mass degeneracy between the molecular states $D^0D_s^{*-}$ and $D^{*0}D_s^-$ indicates that the states may well be mixed. Namely the physical states probably are the mixed states between $D^0D_s^{*-}$ and $D^{*0}D_s^-$. It is necessary to introduce a parameter $\theta$ to describe the mixability for a better understanding the decay properties. In addition, we also investigate the decay properties of the molecular state $D^{*0}D_s^{*-}$ and hope to give some useful theoretical reference for the future experimental exploring. Moreover, we give a brief discussion of the strong decay properties of the four-quark system $c\bar{c}s\bar{u}$ in the compact tetraquark scenario. Our theoretical results are presented as follows.

\subsection{the molecular mixture of $D^0D_s^{*-}$ and $D^{*0}D_s^-$}

$\mathbf{A.1~Under~heavy~quark~spin~symmetry}$~Considering the mass degeneracy between the molecular states $D^0D_s^{*-}$ and $D^{*0}D_s^-$, we treat the physical states as the mixed states between $D^0D_s^{*-}$ and $D^{*0}D_s^-$, i.e.,
\begin{equation}\label{mixd}
\left(\begin{array}{c}|Z_{cs}\rangle_1\cr   |Z_{cs}\rangle_2
\end{array}\right)=\left(\begin{array}{cc} \cos\theta & \sin\theta \cr -\sin\theta &\cos\theta
\end{array}\right)
\left(\begin{array}{c} D^0D_s^{*-} \cr D^{*0}D_s^-
\end{array}\right).
\end{equation}
The mixing angle $\theta$ can vary in the range of $(0^{\circ}\sim180^{\circ})$. Thus, taking the mixing angle $\theta=45^{\circ}$ we obtain that
\begin{eqnarray}\label{mixing2}
|Z_{cs}\rangle_1^{\theta=45^{\circ}}&=&\frac{1}{\sqrt{2}}(D^0D_s^{*-}+D^{*0}D_s^{-}),\nonumber\\
|Z_{cs}\rangle_2^{\theta=45^{\circ}}&=&\frac{1}{\sqrt{2}}(-D^0D_s^{*-}+D^{*0}D_s^{-}).
\end{eqnarray}

We notice that for the systems with heavy flavour quarks, proper consideration of the heavy quark spin symmetry is necessary. Thus, the $D^0D_s^{*-}$ and $D^{*0}D_s^-$ systems are factorizable in the product of two parts: pure heavy and light flavour quark systems.
That reads
\begin{eqnarray}
D^0D_s^{*-}=-\frac{1}{2}|1_H\otimes 0_L\rangle+\frac{1}{2}|0_H\otimes 1_L\rangle+\frac{1}{\sqrt{2}}|1_H\otimes 1_L\rangle,\nonumber\\
D^{*0}D_s^{-}=-\frac{1}{2}|1_H\otimes 0_L\rangle+\frac{1}{2}|0_H\otimes 1_L\rangle-\frac{1}{\sqrt{2}}|1_H\otimes 1_L\rangle.
\end{eqnarray}
Here, the number 1 and 0 in equations with the subscripts H or L denote the spin of the regrouped daughter heavy or light flavour quark systems.

Combining with Eq.~(\ref{mixing2}), we further obtain that
\begin{eqnarray}\label{mixing3}
|Z_{cs}\rangle_1^{\theta=45^{\circ}}&=&\frac{1}{\sqrt{2}}(-|1_H\otimes 0_L\rangle+|0_H\otimes 1_L\rangle),\nonumber\\
|Z_{cs}\rangle_2^{\theta=45^{\circ}}&=&-|1_H\otimes 1_L\rangle.
\end{eqnarray}
It should be pointed out that each direct product decomposition corresponds to a possible decay mode. The corresponding relations are listed as follows:
\begin{eqnarray}
|1_H\otimes 0_L\rangle &\leftrightarrow& J/\psi K,\nonumber\\
|0_H\otimes 1_L\rangle &\leftrightarrow& \eta_c K^*,\nonumber\\
|1_H\otimes 1_L\rangle &\leftrightarrow& J/\psi K^*.
\end{eqnarray}
The above correspondences indicate that based on heavy quark spin symmetry the mixed molecular state $|Z_{cs}\rangle_1^{\theta=45^{\circ}}$ can decay into the $J/\psi K$ and $\eta_c K^*$ hidden-charm final channels, while its partner state $|Z_{cs}\rangle_2^{\theta=45^{\circ}}$ can't decay via these two decay modes. Meanwhile, the partial width ratio between $J/\psi K$ and $\eta_c K^*$ for $|Z_{cs}\rangle_1^{\theta=45^{\circ}}$ is
\begin{eqnarray}
\frac{\Gamma[|Z_{cs}\rangle_1^{\theta=45^{\circ}}\rightarrow J/\psi K]}{\Gamma[|Z_{cs}\rangle_1^{\theta=45^{\circ}}\rightarrow \eta_c K^*]}=1.
\end{eqnarray}

According to the experimental data, the masses of the newly observed states $Z_{cs}(3985)$ and $Z_{cs}(4000)$ are comparable. Moreover, the $Z_{cs}(4000)$ state is observed in the $J/\psi K$ decay channel while non-observation of the $Z_{cs}(3985)$ signal in
this decay mode. Thus, if the two newly observed $Z_{cs}$ states are two different states, our theoretical results under the heavy quark spin symmetry imply that the molecular state $|Z_{cs}\rangle_1^{\theta=45^{\circ}}$ may be the candidate of the $Z_{cs}(4000)$ state and its partner state $|Z_{cs}\rangle_2^{\theta=45^{\circ}}$ may correspond to the $Z_{cs}(3985)$ state. In addition, if the mixed state $|Z_{cs}\rangle_1^{\theta=45^{\circ}}$ is the newly observed state $Z_{cs}(4000)$ indeed, except for the $J/\psi K$ decay mode the future experiments may find this state in the $\eta_c K^*$ final channel as well.

$\mathbf{A.2~With~the~quark-exchange~model}$~Based on heavy quark spin symmetry we analyse the decay properties of the molecular mixtures $|Z_{cs}\rangle_1^{\theta=45^{\circ}}$ and $|Z_{cs}\rangle_2^{\theta=45^{\circ}}$. Our results imply that the newly observed states $Z_{cs}(3985)$ and $Z_{cs}(4000)$ may be explained as $\frac{1}{\sqrt{2}}(-D^0D_s^{*-}+D^{*0}D_s^{-})$ and $\frac{1}{\sqrt{2}}(D^0D_s^{*-}+D^{*0}D_s^{-})$, respectively. Considering the heavy quark spin symmetry is only an approximation, more reliable calculations are desperately needed. Hence we further explore the decay properties of the molecular mixtures in the framework of the quark-exchange model.

\begin{figure}[!t]
\centering \includegraphics[height=8cm,width=6cm]{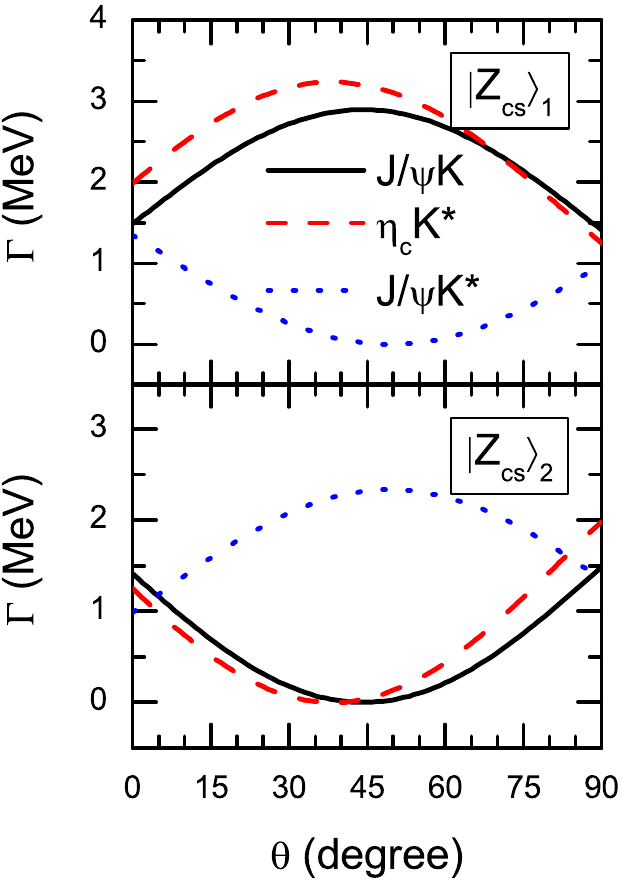} \caption{The partial decay widths (MeV) for the mixed states $|Z_{cs}\rangle_1$ and $|Z_{cs}\rangle_2$ as a function of the mixing angle. The masses and root mean square radius of the two mixed states are
fixed on 4003 MeV and $r_{mean}=1.0$ fm, respectively.}\label{widths1}
\end{figure}

\begin{table*}[!]
\centering\caption{\label{decaywidths} The partial decay widths (MeV) for the mixed states $|Z_{cs}\rangle_1$ and $|Z_{cs}\rangle_2$ with the mixing angle and root mean square radius fixed at $\theta=45^{\circ}$ and $r_{mean}=1.0$ fm, respectively.}
\begin{tabular}{ccccccccccccccc}\hline\hline
state ~~~ &mass(MeV) ~~~  &$\Gamma[J/\psi K]$ ~~~ &$\Gamma[\eta_c K^*]$ ~~~ &$\Gamma[J/\psi K^*]$~~~ &observed~state\\
\hline
$|Z_{cs}\rangle_1^{\theta=45^{\circ}}$~~~&4003 ~~~ &2.89~~~&3.19~~~&0.01 ~~~&$Z_{cs}(4000)$\\
$|Z_{cs}\rangle_2^{\theta=45^{\circ}}$~~~&3982 ~~~ &0.01~~~&0.04~~~&-~~~ &$Z_{cs}(3985)$\\
\hline\hline
\end{tabular}
\end{table*}

According to the mixing scheme defined in Eq.~(\ref{mixd}), in Fig.~\ref{widths1} we plot the hidden-charm decay widths of the mixed states $|Z_{cs}\rangle_1$ and $|Z_{cs}\rangle_2$ as a function of the mixing angle $\theta$ in the region of ($0^{\circ}\sim90^{\circ}$) by fixing the masses of the two states at $M=4003$ MeV and root mean square radius at $r_{mean}=1.0$ fm. From the figure, we get that the variation curves like a bowel structure when the mixing angle varies from $\theta=0^{\circ}$ to $\theta=90^{\circ}$, and the changing trends of the decay modes $J/\psi K$ and $\eta_c K^*$ are opposite to that of the decay mode $J/\psi K^*$. Meanwhile, we notice that with the mixing angle increasing to $\theta\simeq45^{\circ}$, the partial decay widths of the $J/\psi K$ and $\eta_c K^*$ modes for the mixed state $|Z_{cs}\rangle_1$ can reach a maximum about $\Gamma\sim3$ MeV, while that for the mixed state $|Z_{cs}\rangle_2$ decrease into the minimum about $\Gamma\sim0$ MeV. The above decay properties of the mixed states $|Z_{cs}\rangle_1$ and $|Z_{cs}\rangle_2$ obtained here are consistent with those gotten under heavy quark spin symmetry.

\begin{figure}[!t]
\centering \includegraphics[height=8cm,width=6cm]{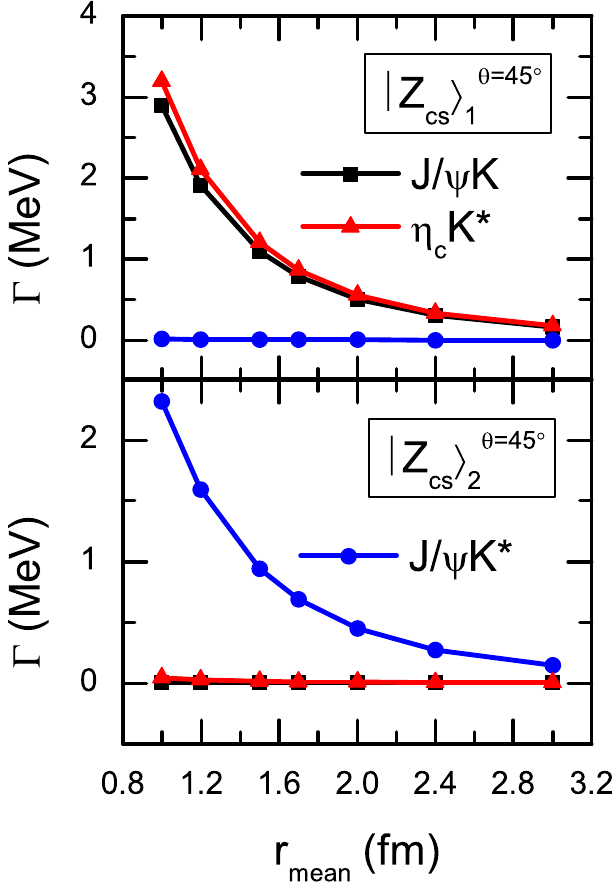} \caption{The partial decay widths (MeV) for the mixed states $|Z_{cs}\rangle_1^{\theta=45^{\circ}}$ and $|Z_{cs}\rangle_2^{\theta=45^{\circ}}$ as a function of the root mean square radius $r_{mean}$. The masses of the two mixed states are both
fixed on 4003 MeV.}\label{widths2}
\end{figure}

Fixing the mixing angle at $\theta=45^{\circ}$(see Table~\ref{decaywidths} ), we obtain
\begin{eqnarray}
\Gamma[|Z_{cs}\rangle_1^{\theta=45^{\circ}}\rightarrow J/\psi K]&\simeq&2.89~\text{MeV},\nonumber\\
\Gamma[|Z_{cs}\rangle_1^{\theta=45^{\circ}}\rightarrow \eta_c K^*]&\simeq&3.19~\text{MeV},
\end{eqnarray}
which are large enough to be observed in experiments. Observation of the state $Z_{cs}(4000)$ in the $J/\psi K$ decay mode indicates $Z_{cs}(4000)$ may be a candidate of the mixed state $|Z_{cs}\rangle_1^{\theta=45^{\circ}}$. Meanwhile, the partial decay widths ratio between $J/\psi K$ and $\eta_c K^*$ is
\begin{eqnarray}
\frac{\Gamma[|Z_{cs}\rangle_1^{\theta=45^{\circ}}\rightarrow J/\psi K]}{\Gamma[|Z_{cs}\rangle_1^{\theta=45^{\circ}}\rightarrow \eta_c K^*]}=0.9,
\end{eqnarray}
which is close to unit. Namely if the observed state $Z_{cs}(4000)$ can be explained as the mixed state $|Z_{cs}\rangle_1^{\theta=45^{\circ}}$, the mode $\eta_c K^*$ may be a ideal channel as well to further decode its inner structure.

Then, we investigate the decay properties of its partner state $|Z_{cs}\rangle_2^{\theta=45^{\circ}}$ and collect in Table~\ref{decaywidths} as well. From the table, it is obvious that for the state $|Z_{cs}\rangle_2^{\theta=45^{\circ}}$, the partial decay width of $J/\psi K$ is small to zero which agrees with the decay properties of the newly observed state $Z_{cs}(3985)$: non-observation in the $J/\psi K$ mode. Thus, we take the mixed state $|Z_{cs}\rangle_2^{\theta=45^{\circ}}$ as the state $Z_{cs}(3985)$, and fix the mass on $M=3982$ MeV. We obtain that the decay width of the mode $\eta_c K^*$ is small to zero as well. Meanwhile, the mode $J/\psi K^*$ is forbidden for the mass below the threshold of $J/\psi K^*$.

It should be mentioned that all of the above theoretical predictions are obtained with the root mean square radius $r_{mean}=1.0$ fm. However, the
root mean square radius is not determined absolutely, which bares a large uncertainty. Thus fixing the masses of the mixed states $|Z_{cs}\rangle_1^{\theta=45^{\circ}}$ and $|Z_{cs}\rangle_2^{\theta=45^{\circ}}$ at $M=4003$ MeV, we further explore the decay properties as a function of the root mean square radius $r_{mean}$. The results are shown in Fig.~\ref{widths2}. One notes that the bigger $r_{mean}$ value leads to a narrower decay width. A loose understanding is that with the $r_{mean}$ increasing the two mesons($D^{(*)0}$ and $D_s^{(*)-}$) in the four-quark state farther separate from each other, thus, the charm and anti-charm quarks are more difficult to form a charmonium.

In addition, we notice that as the consequence of $Z_{cs}(4000)$ and $Z_{cs}(3985)$ being most pure $|Z_{cs}\rangle_1^{\theta=45^{\circ}}$ and $|Z_{cs}\rangle_2^{\theta=45^{\circ}}$ states, respectively, the branching ratio $\frac{\Gamma[Z_{cs}\rightarrow D^0D_s^{*-}]}{\Gamma[Z_{cs}\rightarrow D^{*0}D_s^-]}\simeq0.5$ for $Z_{cs}(4000)$ or $Z_{cs}(3985)$. This result is consistent with the experimental analysis for the $Z_{cs}(3985)$ state~\cite{Ablikim:2020hsk}.

The big difference of their decay widths leads to intensive discussions on whether $Z_{cs}(3985)$ and $Z_{cs}(4000)$ are two different states and possible implications if they are different states in literature~\cite{Meng:2021rdg}.   In the present work, we adopted the quark-exchange model, which is a suitable framework for the hidden-charm decay modes. Unfortunately, the quark-exchange model is not suitable for the so-called "fall-apart" open-charm decay modes. For the $Z_{cs}$ states, the open-charm decay modes are the dominant decay modes according to BESIII and BELLEII measurement. Therefore, we do not discuss the total widths of the $Z_{cs}$ states in this work.

\subsubsection{The molecular state $D^{*0}D_s^{*-}$}

As mentioned in the previous section, the authors in Ref.~\cite{Meng:2021rdg} suggested a tensor $D^*D_s^{*}$ resonance, and its mass and decay width were predicted to be $M=4126$ MeV and $\Gamma=13$ MeV, respectively. The sizable predicted decay width and advanced experimental methods indicate that this tensor resonance has a good potential to be observed in future experiments by BESIII or LHCb collaborations. In this work we investigate the hidden-charm decay properties of the $D^*D_s^{*}$ molecular states with different spin-parity, and hope to provide reference for future experimental explorations.

$\mathbf{B.1~Under~heavy~quark~spin~symmetry}$~For the molecular state $D^{*0}D_s^{*-}$, the spin-parity($J^P$) has three different values: $0^+$, $1^+$ and $2^+$. Here we label them as $|D^{*0}D_s^{*-}\rangle_{0^+}$, $|D^{*0}D_s^{*-}\rangle_{1^+}$ and $|D^{*0}D_s^{*-}\rangle_{2^+}$, respectively. Similarly, the $D^{*0}D_s^{*-}$ system can be factorized into pure heavy and light flavour quark parts under heavy quark spin symmetry, i.e.,
\begin{eqnarray}\label{eq4}
|D^{*0}D_s^{*-}\rangle_{0^+}&=&\frac{\sqrt{3}}{2}|0_H\otimes 0_L\rangle-\frac{1}{2}|1_H\otimes 1_L\rangle,\nonumber\\
|D^{*0}D_s^{*-}\rangle_{1^+}&=&\frac{1}{\sqrt{2}}|1_H\otimes 0_L\rangle+\frac{1}{\sqrt{2}}|0_H\otimes 1_L\rangle,\nonumber\\
|D^{*0}D_s^{*-}\rangle_{2^+}&=&|1_H\otimes 1_L\rangle.
\end{eqnarray}
Here the direct product decomposition $|0_H\otimes 0_L\rangle$ corresponds to the possible decay mode $\eta_cK$.

From Eq.~(\ref{eq4}), we can see that the molecular state $|D^{*0}D_s^{*-}\rangle_{0^+}$ can decay into the $\eta_cK$ and $J/\psi K^*$ final channels, and the partial decay width ratio between the two channels is
\begin{eqnarray}
\frac{\Gamma[|D^{*0}D_s^{*-}\rangle_{0^+}\rightarrow \eta_cK]}{\Gamma[|D^{*0}D_s^{*-}\rangle_{0^+}\rightarrow J/\psi K^*]}\simeq1.7.
\end{eqnarray}
For the molecular state $|D^{*0}D_s^{*-}\rangle_{1^+}$, the decay channel $J/\psi K^*$ is forbidden and the partial decay width ratio between $J/\psi K$ and $\eta_cK^*$ is
\begin{eqnarray}
\frac{\Gamma[|D^{*0}D_s^{*-}\rangle_{1^+}\rightarrow J/\psi K]}{\Gamma[|D^{*0}D_s^{*-}\rangle_{1^+}\rightarrow \eta_cK^*]}\simeq1.
\end{eqnarray}
As to the molecular state $|D^{*0}D_s^{*-}\rangle_{2^+}$, it can decay into $J/\psi K^*$ only in all hidden-charm strong decay modes.

Of course, the above results are obtained based on heavy quark spin symmetry, which is rather rough for the $c$ quark having a limited mass. While the main predictions should hold and be helpful for future experiments.

\begin{figure}[!t]
\centering \includegraphics[height=10cm,width=10cm]{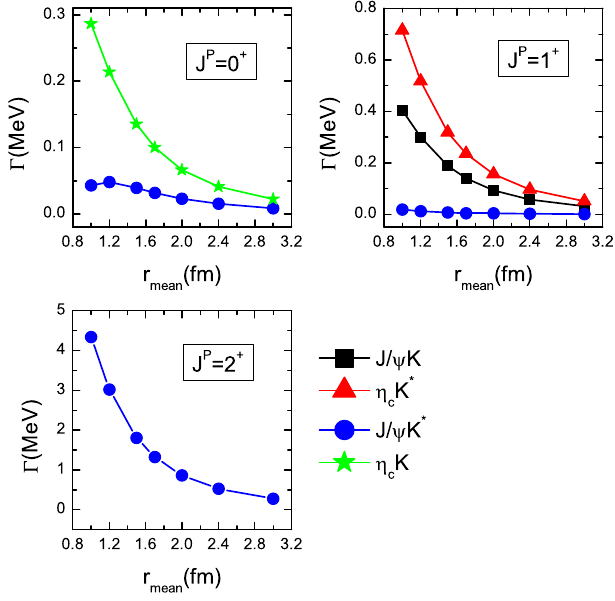} \caption{The partial decay widths (MeV) for the $D^{*0}D_s^{*-}$ molecular states as a function of the root mean square radius $r_{mean}$ with spin-parity $J^P=0^+$, $1^+$ and $2^+$ . The masses
are fixed on 4126 MeV.}\label{widths3}
\end{figure}

$\mathbf{B.2~With~the~quark-exchange~model}$~Furthermore, we analyze the hidden-charm decay properties of the molecular state $D^{*0}D_s^{*-}$ using the quark-exchange model. Fixing the masses at $M=4126$ MeV, we plot the variations of the decay widths as functions of the root mean square radius $r_{mean}$ in Fig.~\ref{widths3}. From the figure, we find that with the $r_{mean}$ increasing in the range of (1.0-3.0) fm, the partial decay widths for $D^{*0}D_s^{*-}$ decaying into the hidden-charm channels decrease. Moreover, the partial decay widths for $|D^{*0}D_s^{*-}\rangle_{0^+}$ and $|D^{*0}D_s^{*-}\rangle_{1^+}$ decaying into the hidden-charm channels are less than $\sim0.8$ MeV with the root mean square radius $r_{mean}$ varied in the whole range we considered in the present work, while the $J/\psi K^*$ decay width of the $|D^{*0}D_s^{*-}\rangle_{2^+}$ molecular state can reach up to a few MeV.

\begin{table}[h]
\centering\caption{\label{decaywidthme} The partial decay widths (MeV) for the molecular state $D^{*0}D_s^{*-}$ with the mass and root mean square radius fixed at $M=4126$ MeV and $r_{mean}=1.5$ fm, respectively.}
\begin{tabular}{ccccccccccccccc}\hline\hline
state ~~~ &$\Gamma[\eta_c K]$ ~~~ &$\Gamma[J/\psi K]$ ~~~ &$\Gamma[\eta_c K^*]$ ~~~ &$\Gamma[J/\psi K^*]$\\
\hline
$|D^{*0}D_s^{*-}\rangle_{0^+}$   ~~~  &0.14 ~~~  &- ~~~  &-  ~~~  &0.04\\
$|D^{*0}D_s^{*-}\rangle_{1^+}$  ~~~  &- ~~~  &0.19 ~~~ &0.32 ~~~  &0.01\\
$|D^{*0}D_s^{*-}\rangle_{2^+}$   ~~~  &- ~~~  &- ~~~ &- ~~~  &1.81\\
\hline\hline
\end{tabular}
\end{table}

Fixing the root mean square radius at $r_{mean}=1.5$ fm, we collect the predicted decay properties in Table~\ref{decaywidthme}. As show in table, we can see that the $J/\psi K^*$ decay width of the molecular state $|D^{*0}D_s^{*-}\rangle_{1^+}$ is small to zero, which is consistent with the prediction within heavy quark spin symmetry. The $J/\psi K^*$ decay width of the molecular state $|D^{*0}D_s^{*-}\rangle_{0^+}$ is small as well, which is much smaller than that of the molecular state $|D^{*0}D_s^{*-}\rangle_{2^+}$. The ratio is
\begin{eqnarray}
\frac{\Gamma[|D^{*0}D_s^{*-}\rangle_{2^+}\rightarrow J/\psi K^*]}{\Gamma[|D^{*0}D_s^{*-}\rangle_{0^+}\rightarrow J/\psi K^*]}\simeq45.3.
\end{eqnarray}
Thus, the $J/\psi K^*$ decay channel can be used to distinguish those three molecular states from each other. Moreover, the sizable decay width for $|D^{*0}D_s^{*-}\rangle_{2^+}$ decaying into $J/\psi K^*$ indicates that this tensor molecular state has a good potential to be observed in the  $J/\psi K^*$ channel.

\subsection{ The tetraquark scenario }

$\mathbf{C.1~\big[[cs]^{S=0}[\bar{c}\bar{u}]^{S=1}\big]^{S=1}~and~\big[[cs]^{S=1}[\bar{c}\bar{u}]^{S=0}\big]^{S=1}~systems}$ For comparison, we further explore the strong decay properties of the four-quark system $c\bar{c}s\bar{u}$ as a compact tetraquark state. We consider the mixing between $\big[[cs]^{S=0}[\bar{c}\bar{u}]^{S=1}\big]^{S=1}$ and $\big[[cs]^{S=1}[\bar{c}\bar{u}]^{S=0}\big]^{S=1}$,i.e.,
\begin{equation}\label{mixt}
\left(\begin{array}{c}|T_{cs}\rangle_1\cr   |T_{cs}\rangle_2
\end{array}\right)=\left(\begin{array}{cc} \cos\theta & \sin\theta \cr -\sin\theta &\cos\theta
\end{array}\right)
\left(\begin{array}{c} \big[[cs]^{S=0}[\bar{c}\bar{u}]^{S=1}\big]^{S=1} \cr \big[[cs]^{S=1}[\bar{c}\bar{u}]^{S=0}\big]^{S=1}
\end{array}\right).
\end{equation}
Similarly we investigate the decay properties of the mixed tetraquark states $|T_{cs}\rangle_1$ and $|T_{cs}\rangle_2$ as a function of the mixing angle $\theta$ with the masses fixed at $M=4003$ MeV. Our calculations indicate that all the partial decay widths for the two mixed tetraquark states  decaying into $J/\psi K$, $\eta_c K^*$ and $J/\psi K^*$ are less than $\sim$0.1 MeV with the mixing angle in the range of $(0^{\circ}\sim90^{\circ})$.

\begin{table}[h]
\centering\caption{\label{decaywidtht} The partial decay widths (MeV) for the mixed states $|T_{cs}\rangle_1$ and $|T_{cs}\rangle_2$ with the mixing angle fixed at $\theta=45^{\circ}$.}
\begin{tabular}{ccccccccccccccc}\hline\hline
state ~~~~~ &mass ~~~~~ &$\Gamma[J/\psi K]$ ~~~~~ &$\Gamma[\eta_c K^*]$ ~~~~~ &$\Gamma[J/\psi K^*]$\\
\hline
$|T_{cs}\rangle_1^{\theta=45^{\circ}}$   ~~~~~  &4003 ~~~~~  &0.07 ~~~~~  &0.08  ~~~~~  &0.01\\
$|T_{cs}\rangle_2^{\theta=45^{\circ}}$  ~~~~~  &3982 ~~~~~  &0.00 ~~~~~ &0.02 ~~~~~  &-\\
\hline\hline
\end{tabular}
\end{table}

In the same way, we fix the mixing angle at $\theta=45^{\circ}$ and collected the decay properties in Table~\ref{decaywidtht}. From the table, it is seen that the decay properties of the mixed states $|T_{cs}\rangle_1^{\theta=45^{\circ}}$ and $|T_{cs}\rangle_2^{\theta=45^{\circ}}$ are inconsistent with that of the newly observed $Z_{cs}(4000)$ state. Thus, the possibility of the mixed tetraquark states as the candidates of the newly observed $Z_{cs}$ states may be excluded.

$\mathbf{C.2~\big[[cs]^{S=1}[\bar{c}\bar{u}]^{S=1}\big]^{S=0,1,2}~systems}$  In addition, we also systematically study the hidden-charm decay properties of the compact tetraquark states $\big[[cs]^{S=1}[\bar{c}\bar{u}]^{S=1}\big]^{S=0,1,2}$. Fixing the masses at $M=4126$ MeV, we collect their decay properties in Table~\ref{decaywidtht2}.

\begin{table}[h]
\centering\caption{\label{decaywidtht2} The partial decay widths (MeV) for the compact tetraquark states $\big[[cs]^{S=1}[\bar{c}\bar{u}]^{S=1}\big]^{S=0,1,2}$ with their masses fixed at $M=4126$ MeV.}
\begin{tabular}{ccccccccccccccc}\hline\hline
state ~ &$\Gamma[\eta_c K]$ ~ &$\Gamma[J/\psi K]$ ~ &$\Gamma[\eta_c K^*]$ ~ &$\Gamma[J/\psi K^*]$\\
\hline
$\big[[cs]^{S=1}[\bar{c}\bar{u}]^{S=1}\big]^{S=0}$  ~  &0.13 ~  &- ~  &- ~  &0.77\\
$\big[[cs]^{S=1}[\bar{c}\bar{u}]^{S=1}\big]^{S=1}$  ~  &- ~  &0.11 ~ &0.18 ~  &0.01\\
$\big[[cs]^{S=1}[\bar{c}\bar{u}]^{S=1}\big]^{S=2}$  ~  &- ~  &- ~ &- ~  &0.02\\
\hline\hline
\end{tabular}
\end{table}
According to the table, it is found that in the compact tetraquark scenario the partial decay widths are rather small($\Gamma<0.2$ MeV) except for
$\big[[cs]^{S=1}[\bar{c}\bar{u}]^{S=1}\big]^{S=0}$ decaying into $J/\psi K^*$, which is about $\Gamma\sim0.8$ MeV. Moreover,
\begin{eqnarray}
\frac{\Gamma[\big[[cs]^{S=1}[\bar{c}\bar{u}]^{S=1}\big]^{S=2}\rightarrow J/\psi K^*]}{\Gamma[\big[[cs]^{S=1}[\bar{c}\bar{u}]^{S=1}\big]^{S=0}\rightarrow J/\psi K^*]}\simeq0.03.
\end{eqnarray}
Thus, the compact tetraquark state $\big[[cs]^{S=1}[\bar{c}\bar{u}]^{S=1}\big]^{S=0}$ may be found in the $J/\psi K^*$ channel in future experiments.

\section{Summary}\label{summary}

In the present work we explore the partial decay widths of the $\eta_c K$, $J/\psi K$, $\eta_c K^*$ and $J/\psi K^*$ modes for the four-quark system $c\bar{c}s\bar{u}$ in the molecular and compact tetraquark scenrios using the quark-exchange model. Our main theoretical results are listed as follows.

In the molecular scenario, we systematically investigate the decay properties of the molecular states $D^{(*)0}D_s^{(*)-}$. For the mixed molecular states between $D^{0}D_s^{*-}$ and $D^{*0}D_s^{-}$, we get that the $J/\psi K$ decay width of the mixture $|Z_{cs}\rangle_1^{\theta=45^{\circ}}$($\frac{1}{\sqrt{2}}(D^0D_s^{*-}+D^{*0}D_s^{-})$) is about $\Gamma[|Z_{cs}\rangle_1^{\theta=45^{\circ}}\rightarrow J/\psi K]\sim2.89$ MeV, while that of the mixture $|Z_{cs}\rangle_2^{\theta=45^{\circ}}$($\frac{1}{\sqrt{2}}(-D^0D_s^{*-}+D^{*0}D_s^{-})$) is small to zero. Considering $Z_{cs}(4000)$ observed in the $J/\psi K$ and non-observation of the $Z_{cs(3985)}$ signal in this decay mode, if those two newly observed states are two different states, $Z_{cs}(4000)$ may be interpreted as the mixture $|Z_{cs}\rangle_1^{\theta=45^{\circ}}$, while $Z_{cs}(3985)$ may correspond to the mixture $|Z_{cs}\rangle_2^{\theta=45^{\circ}}$. In addition, if the state $Z_{cs}(4000)$ can be explained as the mixed state $|Z_{cs}\rangle_1^{\theta=45^{\circ}}$, the partial decay width ratio between $J/\psi K$ and $\eta_cK^*$ is close to unit, which indicates the decay channel $\eta_cK^*$ may be a ideal channel as well to decode the inner structure of $Z_{cs}(4000)$.

For the molecular state $D^{*0}D_s^{*-}$, our results imply that the partial decay width for $|D^{*0}D_s^{*-}\rangle_{2^+}$ decaying into $J/\psi K^*$ can reach up to $\sim2$ MeV, which is large enough to be observed in experiments.

In the tetraquark scenario, we consider the mixing between $\big[[cs]^{S=0}[\bar{c}\bar{u}]^{S=1}\big]^{S=1}$ and $\big[[cs]^{S=1}[\bar{c}\bar{u}]^{S=0}\big]^{S=1}$ as well. Our calculations indicate that all the partial hidden-charm decay widths are less than $\sim0.1$ MeV with the mixing angle in the range of ($0^{\circ}\sim90^{\circ}$).  Thus, the possibility of the mixed compact tetraquark states as the candidates of the newly observed $Z_{cs}$ states may be excluded.

For the compact tetraquark states $\big[[cs]^{S=1}[\bar{c}\bar{u}]^{S=1}\big]^{S=0,1,2}$, Except the partial decay width for $\big[[cs]^{S=1}[\bar{c}\bar{u}]^{S=1}\big]^{S=0}$ decaying into $J/\psi K^*$ being sizable($\sim0.8$ MeV), the rest of the partial decay widths are small($<0.2$ MeV).

\section*{Acknowledgements }

We would like to thank Shi-Lin Zhu and Guang-Juan Wang for very helpful discussions. This work is supported by the National Natural
Science Foundation of China under Grants No.12005013, No.11947048.

 % \bibliography{document}

\end{document}